\documentclass[conference,a4paper]{IEEEtran}

\usepackage{amsmath}
\usepackage[T1]{fontenc}
\usepackage{newtxtext,newtxmath}
\usepackage{graphicx}
\usepackage{booktabs}
\usepackage{url}
\usepackage[numbers,sort]{natbib}
\usepackage[hidelinks]{hyperref}
\newcommand{\PaperTitle}{Zero-shot diffusion image compression with sectioned codebooks and one channel code}
\newcommand{\dB}{\,dB}
\newcommand{\bpp}{\,bpp}
\ExplSyntaxOn
\cs_new_protected:Npn \ReadReferenceList #1
{
  \file_if_exist:nTF {#1}
  {
    \file_get:nnN {#1} {} \l_tmpa_tl
    \regex_replace_all:nnN
      {(\d) -{2,} (\d)}
      {\1 \c{nobreakspace} to \c{nobreakspace} \2}
      \l_tmpa_tl
    \tl_use:N \l_tmpa_tl
  }
  {\iow_term:x {No~file~#1.}}
}
\ExplSyntaxOff
\makeatletter
\let\PlainBibliography\bibliography
\renewcommand{\bibliography}[1]{%
  \begingroup
  \let\@input@\ReadReferenceList
  \PlainBibliography{#1}%
  \endgroup
}
\makeatother
\hypersetup{pdftitle={\PaperTitle},pdfauthor={Dinanath Padhya}}

\newcommand{\appref}[1]{Appendix~\ref{#1}}
\newcommand{\TableText}{\footnotesize}
\newenvironment{widefigure}{\begin{figure*}[!tp]\setkeys{Gin}{trim=3pt 3pt 3pt 3pt}}{\end{figure*}}
\newenvironment{widetable}{\begin{table*}[!tp]}{\end{table*}}
\newenvironment{landscapetable}{\begin{table*}[!tp]}{\end{table*}}
\makeatletter
\renewcommand{\abstract}{\normalfont\@IEEEabskeysecsize\bfseries\textit{\abstractname}:\hspace{0.67em}\relax
  \@IEEEgobbleleadPARNLSP}
\renewcommand{\IEEEkeywords}{\normalfont\@IEEEabskeysecsize\bfseries\textit{\IEEEkeywordsname}:\hspace{0.67em}\relax
  \@IEEEgobbleleadPARNLSP}
\makeatother

\begin{document}
\title{\PaperTitle}
\author{\IEEEauthorblockN{Dinanath Padhya}
\IEEEauthorblockA{Department of Electronics and Computer Engineering\\
Thapathali Campus, Institute of Engineering, Tribhuvan University\\
Kathmandu, Nepal\\
dinanath@tcioe.edu.np}}
\maketitle

\begin{abstract}
Zero-shot diffusion codecs such as Turbo-DDCM compress images without any
training. In each coded step, the encoder selects noise vectors from one
codebook and transmits a single index for the selected set. We propose
sectioned codebooks. A coded step is divided into sections, and each section
has its own large codebook and transmits its best vector as a short field of
bits. A model based on order statistics quantifies how well the bits of a step
are used and predicts the gain of sectioned codebooks from measurements of
Turbo-DDCM alone. Fast Walsh-Hadamard transforms make sections of $2^{20}$
vectors practical. At equal bits, sectioned codebooks improve the PSNR of
Turbo-DDCM by 0.18 to 0.51\dB{} on three datasets, improve LPIPS and DISTS,
and save 16 to 21\% of the bits at equal PSNR. Codecs that are robust to
channel errors by design are usually compared only with unprotected codecs. We
compare them, at equal total bits, with a codec protected by one channel code.
With one BCH code, the sectioned codec outperforms all reported results of
Robust Turbo-DDCM and BitFlat on two datasets, by up to 7.74\dB{} at the
highest bit error rate. At the lowest bit error rate its margin over BitFlat
is only 0.08 to 0.16\dB. When the code fails, the receiver can still decode
the fields as received. Additional experiments cover burst errors and packet
loss, and all observed gains of robust codecs stay within a limit derived from
the channel capacity.

\end{abstract}


\section{Introduction}
\label{sec:introduction}

Zero-shot diffusion codecs compress images with a pretrained diffusion model
and need no training of their own
\cite{theis2022lossy,vonderfecht2025lossy,ohayon2025compressed}. The decoder
runs the reverse diffusion process, and the encoder controls this process
through the noise that is added in each step. In denoising diffusion codebook
models (DDCM) \cite{ohayon2025compressed}, both sides generate the same random
codebook from a shared seed, and the bitstream identifies the noise vectors
that the encoder has selected. Turbo-DDCM \cite{vaisman2026turbo} is a fast
codec of this family. In each coded step it combines about a hundred vectors
of a Gaussian codebook with 16\,384 entries and transmits the selected set as
a single index, which we call the joint index.

It is not known how efficiently this bitstream uses its bits. Turbo-DDCM keeps
the codebook at the size of the latent, and its authors list a theoretical
account of the method as an open problem \cite{vaisman2026turbo}. The
bitstream is also fragile, because a single flipped bit of the joint index
changes most of the selected vectors of a step. Recent work redesigns the
bitstream so that it tolerates bit errors. Robust Turbo-DDCM
\cite{vaisman2026robustness} writes every selected vector as a separate field,
so that a bit error corrupts only one vector. BitFlat \cite{deng2026bitflat}
assigns every field a small codebook of its own, protects each field with a
short code and discards the vectors whose field is detected as corrupted. For
packet loss, learned codecs are trained with random loss of parts of their
latent representation \cite{sha2025towards,wang2025resicomp,wei2026every}.
These robust codecs are compared with unprotected codecs or with each other.
The classical alternative, an unmodified codec protected by one channel code
at the same total number of bits, is missing from these comparisons, although
robustness by design also costs bits. The separate fields of Robust
Turbo-DDCM, for example, need 41 to 47\% more bits than the joint index for
the same vectors.

In this paper, we replace the joint index of Turbo-DDCM with \emph{sectioned
codebooks} (Fig.~\ref{fig:method}). A coded step is divided into sections.
Each section has its own codebook and contributes the single vector that best
matches the image, which is written as a field of fixed length. To explain why
this helps, we introduce the \emph{steering} of a step, a measure of how
strongly the transmitted noise points towards the part of the image that is
still missing. A model based on order statistics shows that a section steers
more per bit as its codebook grows and that large sections are more efficient
than the joint index. Since a codebook with a million vectors cannot be
stored, we construct the sections from Hadamard matrices, for which a fast
transform computes all correlations of a block. At exactly the bits of
Turbo-DDCM, sections of $2^{20}$ vectors improve the PSNR by 0.18 to 0.51\dB{}
on three datasets, improve LPIPS and DISTS, and reduce the bitrate by 16 to
21\% at equal PSNR. The model predicts these gains from measurements of
Turbo-DDCM alone.

We then protect the sectioned bitstream with one BCH code and compare it with
the robust codecs at equal total bits. The code is designed once for the
highest bit error rate (BER) at which the robust codecs are tested. Even the
original Turbo-DDCM protected in this way outperforms Robust Turbo-DDCM at
every BER from $10^{-4}$ to $10^{-2}$. The sectioned codec with the same code
outperforms all reported results of Robust Turbo-DDCM and BitFlat in PSNR,
LPIPS and DISTS. Its margin over BitFlat is 4.14 to 5.05\dB{} at a BER of
$10^{-2}$ and 0.08 to 0.16\dB{} at $10^{-4}$, where the comparison with
reported values is not conclusive. Even when the code is designed for twice
the highest reported BER, the sectioned codec matches Turbo-DDCM with a code
for the reported range. When the code fails, the receiver can still decode the
fields as received, which a joint index does not allow.

The main contributions of this paper are as follows.
\begin{itemize}
\item We propose sectioned codebooks for zero-shot diffusion compression,
together with a model of the steering of a coded step that predicts their gain
over the joint index.
\item We construct sections of up to $2^{20}$ vectors from Hadamard matrices
and obtain consistent rate-distortion gains over Turbo-DDCM on three datasets.
\item We show that a codec protected by one channel code is a stronger
baseline than the unprotected codecs used so far, and that the sectioned codec
with one BCH code outperforms the reported results of two robust diffusion
codecs.
\end{itemize}
Additional results on burst errors and packet loss, and a limit on the gain
that robustness by design can achieve, are given in the appendices.

\section{Related work}
\label{sec:related}

Zero-shot diffusion codecs differ in how they transmit the noise of a step.
DiffC \cite{theis2022lossy} uses reverse channel coding, and its practical
implementation codes the noise in chunks of coordinates that are searched
separately \cite{vonderfecht2025lossy}. DDCM \cite{ohayon2025compressed}
selects one vector of a Gaussian codebook per step, and Turbo-DDCM
\cite{vaisman2026turbo} combines many vectors per step and codes their set
jointly. Later codecs change the trajectory that the selections follow
\cite{zhang2026trajectory} or distribute the selections over time, space and
channels \cite{cheng2026spatio}. In the latter, the channels of the latent are
split into groups and one codebook vector is selected for each group. Chunks
and channel groups partition the coordinates of the noise, whereas every
vector of a section spans all coordinates and the selected vectors are added.
BitFlat \cite{deng2026bitflat} gives every field a codebook of 512 or 1\,024
vectors in order to confine bit errors. It does not compare this layout with
the joint index on an error-free channel and does not consider larger
codebooks.

Sectioned codebooks have the structure of sparse regression codes. These codes
were introduced for the Gaussian channel \cite{joseph2012least} and achieve
the rate-distortion function of Gaussian sources with codewords that add one
column from each section of a random matrix
\cite{venkataramanan2014minimum,venkataramanan2014efficient,
venkataramanan2019sparse}. Transmitting the positions of the largest entries
of a randomly rotated vector also achieves this function
\cite{no2016rateless}, which is the principle behind the selection in
Turbo-DDCM. Algebraic codebooks in speech coding place one signed pulse in
each of several tracks \cite{salami1998design}. Design matrices built from
Hadamard matrices make the decoders of sparse superposition codes fast
\cite{rush2017capacity,barbier2017approximate}, and randomized Hadamard
transforms serve as fast random projections
\cite{ailon2009fast,tropp2011improved}. Our codec uses a sparse regression
code with signs and equal weights as the noise of a diffusion step. The theory
of these codes addresses Gaussian sources and long blocks and does not cover
this use. Two concurrent preprints are related to parts of our work. H-DDCM
\cite{ronai2026hddcm} replaces the Gaussian codebook of Turbo-DDCM with
randomized Hadamard dictionaries, from which the best vectors are selected and
coded jointly. Reeves and Venkataramanan \cite{reeves2026lossy} bound the
distortion of sparse regression codes with the order statistics that also
appear in our model. Neither work studies sections in a diffusion codec or
their protection by a channel code.

Protecting an image bitstream with error-correcting and erasure codes,
including unequal protection of embedded bitstreams, is well established for
wavelet coders \cite{albanese1996priority,mohr2000unequal,
dumitrescu2004globally,hamzaoui2005optimized}. Robust codecs take a different
route. Learned codecs are trained to tolerate packet loss
\cite{sha2025towards,wang2025resicomp,wei2026every,cheng2024grace}, and the
robust diffusion codecs redesign the bitstream
\cite{vaisman2026robustness,deng2026bitflat}. The separation theorem states
that separate source and channel coding is optimal for a known channel
\cite{cover2006elements}, and the broadcast approach quantifies what is lost
when the channel state is unknown
\cite{shamai2003broadcast,tian2008successive,tajer2021broadcast}. We use the
classical protection as a baseline for robust diffusion codecs, which previous
comparisons do not include.

\section{Proposed method}
\label{sec:sections}

This section reviews the coded step of Turbo-DDCM, introduces a measure of how
well a step uses its bits, and presents sectioned codebooks, their fast
construction and their protection by one channel code. Fig.~\ref{fig:method}
illustrates the difference between Turbo-DDCM and the proposed codec.

\begin{widefigure}
\centering
\includegraphics{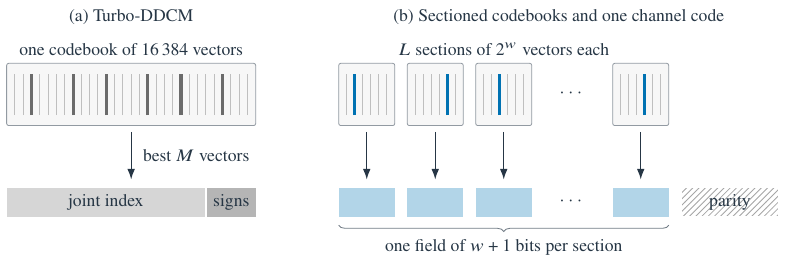}
\caption{One coded step. (a) Turbo-DDCM selects the best $M$ vectors of one
codebook and codes the selected set as a joint index. (b) The proposed codec
selects the best vector of each section and writes one field per section. One
BCH code protects all fields of an image.}
\label{fig:method}
\end{widefigure}

\subsection{Turbo-DDCM}
\label{sec:turbo}

Turbo-DDCM \cite{vaisman2026turbo} uses a pretrained latent diffusion model
\cite{rombach2022high}. The image is mapped to a latent $x_0$ of $D=16\,384$
values, and the reverse process runs for 30 steps from a fixed starting point.
A regular reverse step adds fresh Gaussian noise. A coded step instead adds a
noise vector $z$ that the encoder selects and the decoder reconstructs from
the bitstream. The encoder knows the current prediction $\hat{x}_0$ of the
latent and computes the residual $r=x_0-\hat{x}_0$. Both sides generate a
codebook of $J=D$ Gaussian vectors $c_1,\dots,c_J$ from a shared seed. The
encoder selects the $M$ vectors with the largest correlation magnitude
$|\langle c_j,r\rangle|$, assigns each the sign of its correlation, and sets
$z$ to their sum, scaled to unit variance. The bits of the step are the joint
index, which identifies the selected set among the $\binom{J}{M}$ possible
sets, and the $M$ signs. The last steps of the reverse process carry no bits.

\subsection{Steering of a coded step}
\label{sec:steering}

The transmitted noise is useful through its component along the residual. For
a random codebook, the normalized correlation $g_j=\langle c_j,r\rangle/\lVert
r\rVert$ of a vector is a standard normal variable regardless of the residual.
The signed sum of the selected vectors has length $\sum|g_j|$ along the
residual and a squared length of about $MD$ in the orthogonal directions. This
gives the squared correlation between the noise and the residual,
\begin{equation}
\frac{\langle z,r\rangle^{2}}{\lVert z\rVert^{2}\,\lVert r\rVert^{2}}
=\frac{E}{D+E},\qquad
E=\frac{1}{M}\Bigl(\sum_{\text{selected}}|g_j|\Bigr)^{2}.
\label{eq:steering}
\end{equation}
We call $E$ the \emph{steering} of the step. Random noise has $E\approx1$, and
the coded steps of Turbo-DDCM reach several hundred.

The steering that $b$ bits can achieve is bounded. A description with $b$ bits
cannot approximate the direction of the residual better than the
rate-distortion function of a Gaussian vector allows \cite{cover2006elements},
which gives $E\le D\,(2^{2b/D}-1)$. For the 900 bits per step that serve as an
example in this section, the bound is 1\,296. The joint index is far from this
bound. The $M$ largest of $J$ standard normal magnitudes exceed the threshold
$a$ that a fraction $M/J$ of all magnitudes exceeds, and their mean is
$\bar{g}=2\varphi(a)J/M$, where $\varphi$ is the normal density. The steering
is then $E=M\bar{g}^{2}$. The 91 vectors that Turbo-DDCM transmits with 900
bits give $E=859$, which is 66\% of the bound.

\subsection{Sectioned codebooks}
\label{sec:sectiondesign}

In the proposed codec, a coded step has $L$ sections. Each section is a
separate codebook of $2^{w}$ vectors and contributes the vector with the
largest correlation magnitude, together with the sign of its correlation. The
selection of a section is written as a field of $w+1$ bits, and the bits of
the step are the concatenated fields. The noise is the normalized sum of the
$L$ selected vectors, as in Turbo-DDCM, and the rest of the codec is
unchanged.

From (\ref{eq:steering}), sections of equal size give
\begin{equation}
E=L\,\mu_w^{2},
\label{eq:sections}
\end{equation}
where $\mu_w$ is the expected maximum of $2^{w}$ standard normal magnitudes.
The steering per bit, $\mu_w^{2}/(w+1)$, increases with the section size. It
is 1.08 for $2^{10}$ vectors, 1.14 for $2^{14}$, 1.19 for $2^{20}$ and 1.22
for $2^{24}$, compared with 0.95 for the joint index
(Fig.~\ref{fig:sections}a). Large sections are more efficient, and their size
is limited only by computation. With sections of $2^{20}$ vectors, 900 bits
hold 43 fields and give $E=1\,074$, which is 83\% of the bound. When the bits
of a step are not a multiple of the field length, the sections have two sizes
and we add their contributions $\mu_w^{2}$, which overestimates $E$ by at most
0.5\% for the layouts used here.

Sectioned codebooks differ from Turbo-DDCM in the number of candidate vectors
and in the way the selection is written, and the model separates the two
effects. If the best vectors of the whole pool of all sections were selected
and written as one joint index, each vector would cost about $\log_2
e\approx1.4$ bits more than a field. At equal bits, such an index holds 40
vectors instead of 43 and gives $E=1\,041$. This suggests that most of the
gain over Turbo-DDCM comes from the larger pool and a smaller part from the
fields, which the ablation in Section~\ref{sec:ablation} confirms.

\subsection{Fast sections from Hadamard blocks}
\label{sec:fast}

A section of $2^{20}$ Gaussian vectors has $2^{34}$ values and can be neither
stored nor generated for every step. We therefore construct sections from
blocks of $D$ vectors that need no storage. Let $H$ be the Hadamard matrix of
order $D$, let $\pi$ be a permutation of its columns drawn once per step, and
let $v$ be a vector of $D$ standard normal values drawn once per block. The
vectors of the block are
\begin{equation}
c_j[i]=v[i]\,H[j,\pi(i)],\qquad j=1,\dots,D .
\label{eq:block}
\end{equation}
Each vector has independent standard normal coordinates, as the model assumes.
The correlations of all $D$ vectors with the residual are obtained with one
fast Walsh-Hadamard transform of the values $v[i]\,r[i]$, taken in the order
$\pi$, which costs $D\log_2 D$ additions. A section of $2^{w}\ge D$ vectors
consists of $2^{w}/D$ blocks with independent $v$, and a smaller section uses
the first $2^{w}$ vectors of one block. The decoder needs no transform,
because it regenerates only the $L$ selected vectors from (\ref{eq:block}).

The vectors of a block share the magnitudes of $v$ and are not mutually
independent. This does not affect a section, which contributes a single
vector, and a control experiment with independent Gaussian vectors showed no
significant difference (Section~\ref{sec:ablation}). Dictionaries based on
Hadamard matrices are known from sparse superposition codes
\cite{rush2017capacity,barbier2017approximate} and from fast random
projections \cite{ailon2009fast,tropp2011improved}. The Gaussian values $v$
are specific to our use, in which the sum of the selected vectors replaces
Gaussian noise.

\subsection{Protection with one channel code}
\label{sec:protection}

For transmission over a noisy channel, we protect the whole bitstream of an
image with a single channel code. Of a total of $n$ bits, $k$ bits are used
for the compressed bitstream and the remaining bits for the parity of a
shortened binary BCH code \cite{bose1960class,lin2004error} that corrects up
to $t$ bit errors. A bounded-distance decoder returns the bitstream exactly
when at most $t$ of the transmitted bits are flipped. For independent bit
errors with rate $p$, the probability $P_{\mathrm f}(p)$ that decoding fails
is a binomial tail. If decoding fails, the receiver displays a mid-gray image.
With $Q(k)$ the mean quality of the codec at $k$ bits and $Q_0$ the quality of
the mid-gray image, the expected quality is
\begin{equation}
S(p)=\bigl(1-P_{\mathrm f}(p)\bigr)\,Q(k)+P_{\mathrm f}(p)\,Q_0 .
\label{eq:separate}
\end{equation}
This expression is exact and requires no channel simulation. The code is
designed once for the highest tested BER. Among the operating points and codes
that fit into $n$ bits, we select the pair that maximizes (\ref{eq:separate})
at that BER. \appref{sec:framework} describes the codes in detail.

A sectioned bitstream allows a better reaction to a decoding failure than a
mid-gray image. The code is systematic, and every bit pattern of a field
identifies a valid vector of its section. The receiver can still decode the
fields as received. In this case, a flipped bit replaces the vector of one
section and leaves the other sections unchanged. We refer to this option as
the \emph{fallback}.

\section{Experiments}
\label{sec:experiments}

\subsection{Setup}
\label{sec:setup}

We implemented sectioned codebooks in the public code of Turbo-DDCM
\cite{vaisman2026turbo}, with its pretrained model, weights and seed. Only the
selection of the noise differs from the original codec. Following the robust
codecs \cite{vaisman2026robustness,deng2026bitflat}, we used the 24 Kodak
images \cite{kodak1993suite}, the 100 images of the DIV2K validation set
\cite{agustsson2017ntire} and the 250 images of the CLIC 2020 professional
test set \cite{clic2020test}, each cropped to its central $512\times512$
pixels. Quality is measured with PSNR, LPIPS \cite{zhang2018unreasonable} and
DISTS \cite{ding2022image}. Unless stated otherwise, the codecs use 24 coded
steps, as the robust codecs do, and sections of $2^{20}$ vectors. The
comparisons with Turbo-DDCM are made per image at exactly the same number of
bits, and differences are reported with the standard error of their mean over
the images.

For bit errors, we compare with the results reported for Robust Turbo-DDCM
\cite{vaisman2026robustness} and BitFlat \cite{deng2026bitflat} at BERs from
$10^{-4}$ to $10^{-2}$. BitFlat was reported with one parity bit per field,
which is its main configuration, and with a Hamming or SECDED code per field.
Since the code of BitFlat is not public, we take its results from Tables~I
and~IV of \cite{deng2026bitflat}. To check that our evaluation matches the
reported one, we reproduced Robust Turbo-DDCM, whose code is public. Our PSNR
values are within 0.15\dB{} of the reported ones on Kodak and within 0.09\dB{}
on DIV2K (\appref{sec:flipschecks}). The reported values for CLIC could not be
reproduced, and on that dataset we compare only with our reproduction of
Robust Turbo-DDCM. The BCH codes are designed for a BER of $10^{-2}$, the
highest tested rate, and the total number of bits is that of the compared
robust codec.

\subsection{Compression performance}
\label{sec:compression}

\begin{widetable}
\centering
\caption{Sectioned codebooks with $2^{20}$ vectors per section (Ours) and
Turbo-DDCM (Base) at equal bits. The PSNR gain is given with its standard
error. The last column is the number of images with a higher PSNR.}
\label{tab:compression}
\TableText
\setlength{\tabcolsep}{3.8pt}
\begin{tabular}{@{}rrrrrrrrrr@{}}
\toprule
 & & \multicolumn{3}{c}{PSNR (dB)} & \multicolumn{2}{c}{LPIPS} & \multicolumn{2}{c}{DISTS} & \\
\cmidrule(lr){3-5}\cmidrule(lr){6-7}\cmidrule(lr){8-9}
bpp & Coded steps & Base & Ours & Gain & Base & Ours & Base & Ours & Images \\
\midrule
\multicolumn{10}{@{}l}{\textit{Kodak, 24 images}} \\
0.0404 & 24 & 23.21 & 23.72 & +0.51 (0.05) & 0.221 & 0.184 & 0.142 & 0.117 & 24 \\
0.0558 & 24 & 24.00 & 24.36 & +0.36 (0.04) & 0.163 & 0.135 & 0.102 & 0.084 & 24 \\
0.0591 & 24 & 24.10 & 24.46 & +0.36 (0.03) & 0.155 & 0.130 & 0.096 & 0.079 & 24 \\
0.0824 & 24 & 24.64 & 24.95 & +0.30 (0.03) & 0.120 & 0.104 & 0.074 & 0.065 & 24 \\
0.0999 & 24 & 24.91 & 25.17 & +0.26 (0.02) & 0.107 & 0.097 & 0.065 & 0.060 & 24 \\
\addlinespace[2pt]
0.0646 & 19 & 24.00 & 24.31 & +0.31 (0.03) & 0.137 & 0.120 & 0.080 & 0.071 & 24 \\
0.1257 & 27 & 25.30 & 25.55 & +0.26 (0.02) & 0.097 & 0.088 & 0.061 & 0.056 & 24 \\
0.1523 & 29 & 25.59 & 25.81 & +0.22 (0.02) & 0.090 & 0.083 & 0.057 & 0.054 & 24 \\
0.1939 & 29 & 25.79 & 25.96 & +0.18 (0.01) & 0.084 & 0.080 & 0.054 & 0.052 & 24 \\
\addlinespace
\multicolumn{10}{@{}l}{\textit{DIV2K, 100 images}} \\
0.0591 & 24 & 23.18 & 23.69 & +0.51 (0.03) & 0.171 & 0.142 & 0.116 & 0.096 & 100 \\
0.0824 & 24 & 23.94 & 24.38 & +0.45 (0.02) & 0.130 & 0.111 & 0.087 & 0.075 & 100 \\
0.0992 & 24 & 24.30 & 24.67 & +0.37 (0.02) & 0.115 & 0.101 & 0.076 & 0.068 & 99 \\
\addlinespace[2pt]
0.1523 & 29 & 25.20 & 25.49 & +0.30 (0.01) & 0.094 & 0.086 & 0.066 & 0.061 & 99 \\
0.1939 & 29 & 25.47 & 25.71 & +0.24 (0.01) & 0.086 & 0.082 & 0.061 & 0.057 & 99 \\
\addlinespace
\multicolumn{10}{@{}l}{\textit{CLIC, 250 images}} \\
0.0591 & 24 & 26.42 & 26.89 & +0.47 (0.02) & 0.119 & 0.101 & 0.089 & 0.075 & 243 \\
0.0824 & 24 & 27.10 & 27.49 & +0.39 (0.01) & 0.093 & 0.082 & 0.070 & 0.062 & 245 \\
\bottomrule
\end{tabular}
\end{widetable}

Table~\ref{tab:compression} compares sectioned codebooks with Turbo-DDCM at
equal bits. Sectioned codebooks improve the PSNR in every row, by 0.18 to
0.51\dB, and they also improve LPIPS and DISTS. At 21\,600 bits per image, the
gain is 0.30\dB{} on Kodak, 0.45\dB{} on DIV2K and 0.39\dB{} on CLIC, and the
PSNR is higher on 24 of 24, 100 of 100 and 245 of 250 images. The gain is
largest at low rates and decreases as the rate grows, on Kodak from 0.51\dB{}
at 0.040\bpp{} to 0.18\dB{} at 0.194\bpp, where the rate-distortion curve of
the codec becomes flat. The rows with 19, 27 and 29 coded steps use the number
of steps that Turbo-DDCM selects by default at these rates, and the gain is
preserved (0.18 to 0.31\dB). In terms of rate, the sectioned codec reaches the
PSNR of Turbo-DDCM with 16 to 21\% fewer bits (Fig.~\ref{fig:sections}c).

\begin{widefigure}
\centering
\includegraphics{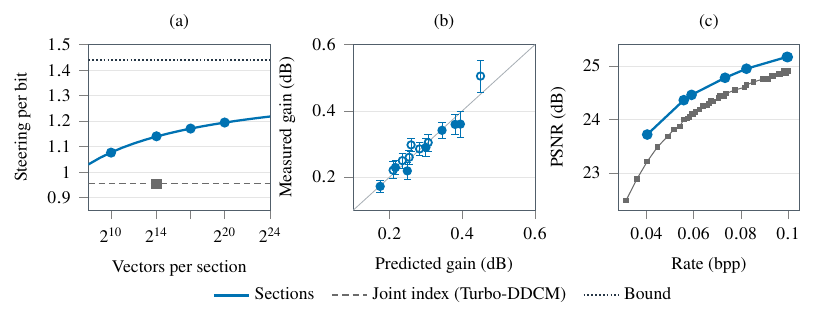}
\caption{Sectioned codebooks on Kodak. (a) Steering per bit as a function of
the section size. Lines show the model and markers show measurements. (b)
Measured and predicted PSNR gain over Turbo-DDCM with standard errors. Open
markers lie outside the range of the fit. (c) Rate-distortion curves with 24
coded steps. Gray is Turbo-DDCM.}
\label{fig:sections}
\end{widefigure}

\subsection{Model validation and ablation}
\label{sec:ablation}

The steering model predicts these gains from measurements of Turbo-DDCM alone.
We fitted the PSNR of Turbo-DDCM on Kodak as a quadratic function of the
logarithm of its steering, using its 32 operating points with 24 coded steps,
and read the expected gain of a sectioned layout from this curve at the
steering given by (\ref{eq:sections}). The first eight predictions were made
before the sectioned codec was run. For the 14 layouts measured on Kodak, with
sections of $2^{10}$ to $2^{20}$ vectors, the measured gain differs from the
prediction by at most 0.05\dB{} (Fig.~\ref{fig:sections}b). The gain also
grows with the section size as predicted, from 0.17\dB{} for $2^{10}$ vectors
to 0.30\dB{} for $2^{20}$ vectors at 21\,600 bits.

\begin{widetable}
\centering
\caption{Ablation on Kodak with 24 coded steps at about 900 bits per step.
PSNR differences are relative to Turbo-DDCM and are given with standard
errors. Time is the encoding time per image.}
\label{tab:ablation}
\TableText
\setlength{\tabcolsep}{4pt}
\begin{tabular}{@{}lrrrrrrr@{}}
\toprule
 & Candidate & Vectors & Bits per & \multicolumn{2}{c}{Steering} & PSNR diff. & Time \\
\cmidrule(lr){5-6}
Codebook & vectors & sent & step & model & measured & (dB) & (s) \\
\midrule
\multicolumn{8}{@{}l}{\textit{Joint index over the best vectors of one pool}} \\
Turbo-DDCM & $1.6\times10^{4}$ & 91 & 900 & 859 & 859 &  & 3.7 \\
One Hadamard block & $1.6\times10^{4}$ & 91 & 900 & 859 & 816 & $-$0.17 (0.02) & 3.5 \\
8 blocks & $1.3\times10^{5}$ & 67 & 892 & 931 & 924 & +0.09 (0.01) & 3.5 \\
60 blocks & $9.8\times10^{5}$ & 54 & 892 & 979 & 977 & +0.20 (0.02) & 3.5 \\
2\,696 blocks & $4.4\times10^{7}$ & 40 & 897 & 1\,041 & 1\,040 & +0.26 (0.02) & 5.8 \\
\addlinespace
\multicolumn{8}{@{}l}{\textit{Sectioned codebooks, one vector per section}} \\
$2^{10}$ vectors per section & $8.3\times10^{4}$ & 82 & 900 & 969 & 969 & +0.17 (0.02) & 3.5 \\
$2^{14}$ vectors per section & $9.8\times10^{5}$ & 60 & 900 & 1\,025 & 1\,026 & +0.22 (0.02) & 3.5 \\
$2^{17}$ vectors per section & $6.6\times10^{6}$ & 50 & 900 & 1\,054 & 1\,054 & +0.29 (0.02) & 3.7 \\
$2^{20}$ vectors per section & $4.4\times10^{7}$ & 43 & 900 & 1\,074 & 1\,075 & +0.30 (0.03) & 5.6 \\
\bottomrule
\end{tabular}
\end{widetable}

Table~\ref{tab:ablation} separates the two sources of the gain. We also
recorded the correlation between the transmitted noise and the residual in
every coded step, which gives the steering directly from (\ref{eq:steering}).
For Turbo-DDCM and all sectioned layouts, the measured steering deviates from
the model by only $-$0.2 to +0.1\%. The size of the pool is the main factor.
With the selection rule and the joint index of Turbo-DDCM, pools of 60 and
2\,696 Hadamard blocks gain 0.20 and 0.26\dB. Writing the selection as fields
adds a smaller part. Sections on these two pools gain 0.22 and 0.30\dB, and
the difference to the joint index on the same pool is significant only for the
larger pool, where it is +0.04\dB{} with a standard error of 0.01\dB. The
table also shows that several vectors should not be taken from the same block.
One block with its best 91 vectors is 0.17\dB{} below Turbo-DDCM, because the
vectors of a block share the values $v$ and their sum is no longer Gaussian.
Sections avoid this, because each section has its own blocks and contributes a
single vector. Finally, sections of $2^{10}$ independent Gaussian vectors and
Hadamard-based sections of the same size differ by $-$0.04\dB{} in PSNR with a
standard error of 0.05\dB, which is not significant.

The encoding cost depends on the section size. At 900 bits per step, the
encoder transforms 2\,696 blocks per step with $6.2\times10^{8}$ additions,
while Turbo-DDCM computes a product with a dense codebook of $D\times J$
values, which takes $2.7\times10^{8}$ multiplications. On a laptop GPU,
encoding an image took 5.6\,s with sections of $2^{20}$ vectors, 3.7\,s with
sections of $2^{17}$ vectors and 3.7\,s with Turbo-DDCM. Sections of $2^{17}$
vectors thus keep 0.29 of the 0.30\dB{} at the encoding time of Turbo-DDCM.
The cost of the transforms grows with the number of sections. At 0.194\bpp,
the highest rate tested, encoding with sections of $2^{20}$ vectors took
8.9\,s compared with 3.8\,s. The decoder regenerates only the selected vectors
and needs no transform.

\subsection{Robustness to bit errors}
\label{sec:flips}

\begin{widetable}
\centering
\caption{PSNR (dB) under independent bit errors. The values of the robust
codecs are those reported in \cite{deng2026bitflat}. The BCH-protected codecs
use the smallest number of bits of each group. The best value of each group is
in bold.}
\label{tab:psnr}
\TableText
\setlength{\tabcolsep}{5pt}
\begin{tabular}{@{}lrrrrrr@{}}
\toprule
 & & \multicolumn{5}{c}{Bit error rate} \\
\cmidrule(l){3-7}
Codec & bpp & $10^{-4}$ & $3\!\times\!10^{-4}$ & $10^{-3}$ & $3\!\times\!10^{-3}$ & $10^{-2}$ \\
\midrule
\multicolumn{7}{@{}l}{\textit{Kodak, 24 images}} \\
Robust Turbo-DDCM \cite{vaisman2026robustness} & 0.1003 & 24.10 & 23.70 & 22.51 & 20.76 & 17.73 \\
BitFlat, parity \cite{deng2026bitflat} & 0.0997 & 24.85 & 24.61 & 23.96 & 22.83 & 20.40 \\
BitFlat, Hamming \cite{deng2026bitflat} & 0.1000 & 24.58 & 24.58 & 24.56 & 24.47 & 23.29 \\
BitFlat, SECDED \cite{deng2026bitflat} & 0.1003 & 24.46 & 24.46 & 24.45 & 24.38 & 23.66 \\
Turbo-DDCM + BCH & 0.0997 & 24.64 & 24.64 & 24.64 & 24.64 & 24.62 \\
Sectioned + BCH (ours) & 0.0997 & \textbf{24.95} & \textbf{24.95} & \textbf{24.95} & \textbf{24.95} & \textbf{24.93} \\
\addlinespace[2pt]
Robust Turbo-DDCM \cite{vaisman2026robustness} & 0.0728 & 23.62 & 23.23 & 22.23 & 20.60 & 17.74 \\
BitFlat, parity \cite{deng2026bitflat} & 0.0735 & 24.36 & 24.15 & 23.64 & 22.61 & 20.32 \\
Turbo-DDCM + BCH & 0.0728 & 24.10 & 24.10 & 24.10 & 24.10 & 24.10 \\
Sectioned + BCH (ours) & 0.0728 & \textbf{24.46} & \textbf{24.46} & \textbf{24.46} & \textbf{24.46} & \textbf{24.46} \\
\addlinespace
\multicolumn{7}{@{}l}{\textit{DIV2K, 100 images}} \\
Robust Turbo-DDCM \cite{vaisman2026robustness} & 0.1003 & 23.28 & 22.86 & 21.62 & 19.70 & 16.64 \\
BitFlat, parity \cite{deng2026bitflat} & 0.0997 & 24.22 & 24.00 & 23.33 & 22.04 & 19.31 \\
BitFlat, Hamming \cite{deng2026bitflat} & 0.1000 & 23.84 & 23.84 & 23.82 & 23.71 & 22.64 \\
BitFlat, SECDED \cite{deng2026bitflat} & 0.1003 & 23.70 & 23.70 & 23.69 & 23.63 & 22.91 \\
Turbo-DDCM + BCH & 0.0997 & 23.92 & 23.92 & 23.92 & 23.92 & 23.92 \\
Sectioned + BCH (ours) & 0.0997 & \textbf{24.38} & \textbf{24.38} & \textbf{24.38} & \textbf{24.38} & \textbf{24.36} \\
\addlinespace[2pt]
Robust Turbo-DDCM \cite{vaisman2026robustness} & 0.0728 & 22.62 & 22.25 & 21.15 & 19.43 & 16.51 \\
BitFlat, parity \cite{deng2026bitflat} & 0.0735 & 23.61 & 23.43 & 22.82 & 21.60 & 19.21 \\
Turbo-DDCM + BCH & 0.0728 & 23.18 & 23.18 & 23.18 & 23.18 & 23.17 \\
Sectioned + BCH (ours) & 0.0728 & \textbf{23.69} & \textbf{23.69} & \textbf{23.69} & \textbf{23.69} & \textbf{23.69} \\
\bottomrule
\end{tabular}
\end{widetable}

\begin{widefigure}
\centering
\includegraphics{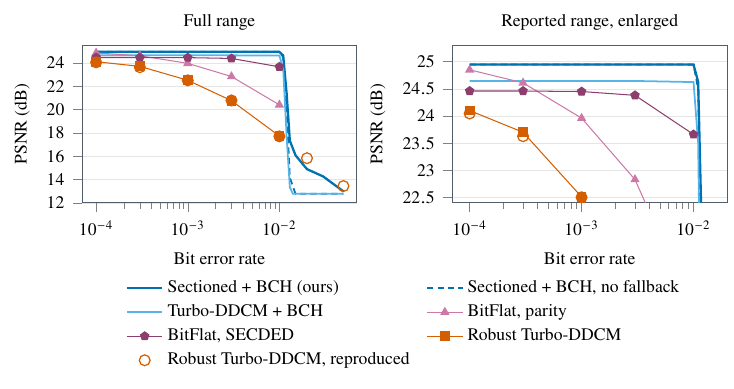}
\caption{PSNR as a function of the BER on Kodak at about 0.10\bpp. The BCH
codes are designed for a BER of $10^{-2}$. Beyond this BER, the sectioned codec may start to fail, and its performance can degrade rapidly.}
\label{fig:flips}
\end{widefigure}

The BCH-protected codecs are compared with the robust codecs in
Table~\ref{tab:psnr} and Fig.~\ref{fig:flips}. At 26\,280 bits per image, the
sectioned codec spends 21\,600 bits on the image and 4\,680 bits on parity,
which corrects up to 319 bit errors. Its quality is constant up to a BER of
$10^{-2}$, whereas the quality of the robust codecs decreases with the BER.

Turbo-DDCM protected by one BCH code already outperforms Robust Turbo-DDCM at
every BER, by 0.48 to 7.29\dB{} on the two datasets. The separate fields of
Robust Turbo-DDCM need so many bits that the joint index can carry more
vectors and their protection within the same total. BitFlat with parity fields
is 0.21 to 0.40\dB{} better than this baseline at a BER of $10^{-4}$ and worse
from a BER of $10^{-3}$ on (\appref{sec:flipsresults}).

The sectioned codec with the same code has the highest PSNR at every BER in
Table~\ref{tab:psnr}. At exactly equal total bits, it outperforms Robust
Turbo-DDCM by 0.84 to 7.74\dB{} and BitFlat with Hamming or SECDED fields by
0.37 to 1.73\dB. Its margin over BitFlat with parity fields is 4.14 to
5.05\dB{} at a BER of $10^{-2}$ and 0.08 to 0.16\dB{} at $10^{-4}$. The margin
at $10^{-4}$ is of the size of the deviations of our reproduction, so the two
codecs should be regarded as equivalent at this BER. LPIPS and DISTS give the
same ranking, and the sectioned codec is better in all 180 comparisons with
reported values (\appref{sec:flipsresults}). Compared with our reproduction of
Robust Turbo-DDCM, it is better by 0.83 to 7.24\dB{} on Kodak, 1.11 to
7.83\dB{} on DIV2K and 1.13 to 9.30\dB{} on CLIC.

The steering model explains the ranking at low BER. The parity fields of
BitFlat are sections of 512 vectors, and their 1\,089 bits per step give
$E=1\,045$. This is more than the 859 of the joint index with the bits that
the code leaves for it, and less than the 1\,074 of sections of $2^{20}$
vectors with the same bits.

\subsection{Beyond the design range}
\label{sec:flipsrange}

A code designed for a BER of $10^{-2}$ fails when the channel is worse. The
BCH-protected Turbo-DDCM then loses the image, while the robust codecs still
deliver a degraded image. The sectioned codec can use the fallback of
Section~\ref{sec:protection}, which we measured on Kodak
(Fig.~\ref{fig:flips}). At a BER of $2\times10^{-2}$, the fallback gives
14.9\dB{} with sections of $2^{20}$ vectors, compared with 12.8\dB{} for a
mid-gray image and 15.8\dB{} for our reproduction of Robust Turbo-DDCM. The
difference to Robust Turbo-DDCM shrinks from $-$3.1 to $-$0.9\dB{} at this BER
and from $-$0.7 to $-$0.5\dB{} at $5\times10^{-2}$. Smaller sections have
shorter fields, so a smaller fraction of the selected vectors is corrupted at
a given BER. With sections of $2^{14}$ vectors, the fallback is on par with
Robust Turbo-DDCM, with differences of 0.0 and +0.2\dB, and with sections of
$2^{10}$ vectors it is above Robust Turbo-DDCM, with differences of +0.9 and
+0.9\dB. Without errors, these smaller sections gain 0.22 and 0.17\dB{} over
Turbo-DDCM instead of 0.30\dB. The section size is thus a trade-off between
quality within the design range and quality beyond it.

The design range can also be extended. A code designed for a BER of
$2\times10^{-2}$, twice the highest reported BER, needs more parity and leaves
736 bits per step for the image at 26\,280 bits. With sectioned codebooks,
this design reaches 24.67\dB{} on Kodak up to a BER of $2\times10^{-2}$. The
difference to Turbo-DDCM with the code designed for $10^{-2}$ is +0.02\dB{} on
Kodak and +0.06\dB{} on DIV2K. In other words, the gain of sectioned codebooks
pays for a design range that is twice as wide. At a BER of $2\times10^{-2}$,
this design is 8.8\dB{} better than our reproduction of Robust Turbo-DDCM on
Kodak. It still outperforms the reported results of Robust Turbo-DDCM and of
BitFlat with Hamming or SECDED fields at every BER, and those of BitFlat with
parity fields from a BER of $10^{-3}$ on, while BitFlat with parity fields is
0.18 to 0.39\dB{} better at $10^{-4}$.

\subsection{Burst errors and packet loss}
\label{sec:other}

We tested two further channels and summarize the results here. When bit errors
occur in bursts, a BCH code designed for independent errors is not suitable,
and we use a Reed-Solomon code instead. With this code, the sectioned codec
outperforms our reproduction of Robust Turbo-DDCM on all 12 burst channels
tested, by 0.39 to 4.30\dB, although the margin becomes small for very long
bursts (\appref{sec:bursts}). For packet loss, we protected the learned codec
MLIC++ \cite{jiang2025mlicpp} with one erasure code and compared it with the
robust codec Every Packet Counts \cite{wei2026every}. The protected codec is
0.14 to 0.92\dB{} better on Kodak and 0.24 to 0.86\dB{} worse on images of two
megapixels, where the comparison is only indicative because the image sets may
differ (\appref{sec:loss}).

\section{Discussion}
\label{sec:discussion}

Sectioned codebooks improve Turbo-DDCM because they draw the noise from a much
larger pool of vectors. The joint index is an efficient description of a set,
but the set is selected from a codebook that is small for the number of bits
spent on it, and the dense codebook of Turbo-DDCM cannot grow. The steering
model also indicates the remaining potential. Sections of $2^{20}$ vectors
reach 83\% of the bound, and larger sections close the gap only slowly. Two
extensions concern the encoder alone. The sections of a step currently choose
simultaneously for the same residual, whereas the encoders of sparse
regression codes let each section react to the choices of the previous ones
\cite{venkataramanan2014efficient}. The model also treats all steps alike,
although the steps of a diffusion process differ in their effect on the image.

The results on error resilience raise the question of when robustness by
design can be better than a channel code. \appref{sec:budget} derives a limit
on this advantage, which we call the budget of a robust codec. A robust codec
cannot deliver more than the reference codec would deliver at the capacity of
the channel. Its possible gain over a code designed for the worst channel
state depends on the range of capacity that the channel states span and on the
slope of the rate-distortion curve. For BERs up to $10^{-2}$, the budget is a
few tenths of a decibel, and all gains of robust codecs that we observed lie
within it. Robustness by design can be expected to pay off when the capacity
range is wide, when the blocks are short, or when no useful bound on the
channel state is known. We therefore suggest that evaluations of robust codecs
include a codec protected by one channel code at equal total bits.

Our study has some limitations as well. BitFlat and Every Packet Counts are not
publicly available, so the comparisons with them rely on reported values. The
sectioned codec was evaluated with one diffusion model, on crops of
$512\times512$ pixels and at rates up to 0.194\bpp. The fallback and the burst
channels were tested on Kodak only. With sections of $2^{20}$ vectors,
encoding is slower than with Turbo-DDCM, and the encoding times come from a
research implementation on a single device. The channel code is a BCH code
with bounded-distance decoding, and a modern code would leave more bits for
the image.

\section{Conclusion}
\label{sec:conclusion}

We proposed sectioned codebooks for zero-shot diffusion image compression.
Each section of a coded step selects its best vector from a large codebook of
its own, and a model based on order statistics predicts the resulting gain
over the joint index of Turbo-DDCM. With sections of $2^{20}$ vectors built
from Hadamard matrices, the PSNR improves by 0.18 to 0.51\dB{} at equal bits
on three datasets, which corresponds to 16 to 21\% fewer bits at equal PSNR.
Protected by one BCH code, the sectioned codec outperforms the reported
results of Robust Turbo-DDCM and BitFlat at all tested BERs, with a clear
margin at high BERs and a marginal one at the lowest. The gain is also
sufficient to double the BER range of the code without falling below
Turbo-DDCM. A codec protected by one channel code is thus a simple and strong
baseline for the evaluation of robust codecs. Future work will study encoders
that choose the sections of a step sequentially and the extension to other
diffusion models and larger images.

\section*{Declarations}

\paragraph{Funding}
This research did not receive any specific grant from funding agencies in the
public, commercial, or not-for-profit sectors.

\paragraph{Competing interests}
The author declares no known competing financial interests or personal
relationships that could have appeared to influence the work reported in this
paper.

\paragraph{Data and code availability}
The implementation and the numerical results of all experiments will be
released with the article. The images and the pretrained codecs are available
from their original distributors under their respective terms.

\section*{Declaration of generative AI and AI-assisted technologies in the manuscript preparation process}
During the preparation of this work the author used Claude to assist with the experimental analysis, interpretation of results, and drafting of the manuscript. The author reviewed and edited the content generated by AI and takes full responsibility for the final content of the manuscript.

\appendices
\section{Comparison rules and channel codes}
\label{sec:framework}

An image is transmitted with a total of $n$ channel bits. The channel has a
state $\theta$ that is unknown to the sender, which is the BER of a binary
channel or the loss rate of a packet channel, and a robust codec is tested
over a range $0\le\theta\le\bar\theta$. A robust codec spends all $n$ bits on
a representation that tolerates errors. The protected codec spends $k$ bits on
the bitstream of an unmodified codec and the remaining bits on the parity of
one channel code.

The comparison follows four rules. The totals are equal, and every transmitted
bit of the protected codec counts, including parity, padding and packet
headers. The images and quality measures are those of the reported results.
The code is designed once for the tested range, and the same bitstream is
evaluated at every channel state. For bit errors, the code is designed for the
highest state $\bar\theta$. For packet loss, where the robust codec is trained
for a uniformly distributed loss rate, the design maximizes the expected
quality under that distribution. Finally, the evaluation is exact, because the
failure probability in (\ref{eq:separate}) is the tail of a known
distribution.

For independent bit errors we use shortened binary BCH codes
\cite{bose1960class,lin2004error}. The primitive narrow-sense BCH code of
length $2^m-1$ that corrects $t$ errors has $r(t)\le mt$ parity bits, given by
the sizes of the cyclotomic cosets of $1,\dots,2t$ modulo $2^m-1$, and
shortening preserves both numbers. For a total of $n$ bits and an operating
point of $k$ bits, $t$ is the largest value with $k+r(t)\le n$, and $m=15$
suffices for the totals used here. The decoding fails when more than $t$ of
the $k+r(t)$ transmitted bits are flipped, so that
\begin{equation}
P_{\mathrm f}(p)=\Pr\bigl[\mathrm{Bin}(k+r(t),p)>t\bigr].
\label{eq:bch}
\end{equation}
A word with more than $t$ errors could also lie within distance $t$ of another
codeword and be accepted as valid. The fraction of such words is below
$2^{-25}$ for all codes used here, and we neglect this case. We verified the
decoding rule by transmitting the bitstreams of all images through an
independent implementation of the codes \cite{hostetter2020galois}. Every word
was recovered exactly when it had at most $t$ errors, and no wrong bitstream
was accepted.

A stronger code than BCH with bounded-distance decoding would leave more bits
for the image. To bound this benefit, we use the converse of Polyanskiy, Poor
and Verd\'u \cite[Theorem 35]{polyanskiy2010channel}. No code with $2^k$
messages and $n$ bits fails less often than one that assigns every message the
$2^{n-k}$ lightest error patterns. We compute this lower bound on the failure
probability exactly and refer to the resulting design as the \emph{limit} of
any code.

For burst errors we use a channel with two states of the Gilbert-Elliott type
\cite{gilbert1960capacity,elliott1963estimates}. The number of flipped bits is
then no longer binomial, and its exact distribution follows from a forward
recursion over the bits. The code for bursts is a Reed-Solomon code
\cite{reed1960polynomial}, which corrects any $t$ corrupted symbols with $2t$
parity symbols. For packet loss, the bitstream is spread over $N$ packets with
a maximum-distance-separable code, so that any $K$ received packets recover it
\cite{rizzo1997effective}. The failure probability for independent loss with
rate $q$ is $\Pr[\mathrm{Bin}(N,1-q)<K]$.

\section{Detailed results for bit errors}
\label{sec:flipsresults}

This appendix describes the reported configurations of the robust codecs and
gives the complete comparison in PSNR, LPIPS and DISTS.

\begin{widefigure}
\centering
\includegraphics{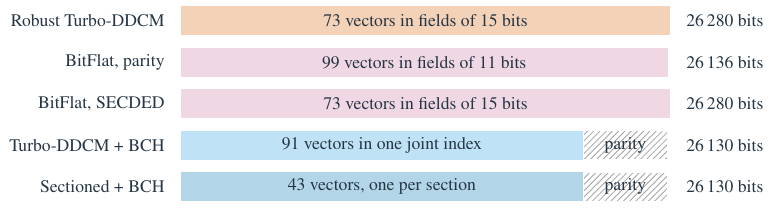}
\caption{Bits of one image with 24 coded steps in five codecs. Lengths are
proportional to bits. The BCH codes are designed for a BER of $10^{-2}$.}
\label{fig:layout}
\end{widefigure}

Fig.~\ref{fig:layout} shows how the compared codecs use the bits of an image.
Robust Turbo-DDCM \cite{vaisman2026robustness} sends each selected vector as a
field of 15 bits. BitFlat \cite{deng2026bitflat} gives every field a codebook
of 512 vectors, which shortens the field, and adds a parity bit, a shortened
Hamming code or a SECDED code to every field. The reported results cover six
configurations with 24 coded steps. These are Robust Turbo-DDCM at 26\,280 and
19\,080 bits per image, BitFlat with parity, Hamming and SECDED fields at
26\,136, 26\,208 and 26\,280 bits, and BitFlat with parity fields at 19\,272
bits. The BERs are $10^{-4}$, $3\times10^{-4}$, $10^{-3}$, $3\times10^{-3}$
and $10^{-2}$.

\begin{landscapetable}
\centering
\caption{Turbo-DDCM with one BCH code minus the reported values of the robust
codecs, at equal total bits. The columns are the BERs in units of $10^{-4}$.
Positive PSNR differences and negative LPIPS and DISTS differences favour the
BCH-protected codec.}
\label{tab:flips}
\TableText
\setlength{\tabcolsep}{3.4pt}
\begin{tabular}{@{}llrrrrrrrrrrrrrrr@{}}
\toprule
 & & \multicolumn{5}{c}{PSNR (dB)} & \multicolumn{5}{c}{LPIPS ($10^{-3}$)}
 & \multicolumn{5}{c}{DISTS ($10^{-3}$)} \\
\cmidrule(lr){3-7}\cmidrule(lr){8-12}\cmidrule(l){13-17}
Robust codec & bpp & 1 & 3 & 10 & 30 & 100 & 1 & 3 & 10 & 30 & 100 & 1 & 3 & 10 & 30 & 100 \\
\midrule
\multicolumn{17}{@{}l}{\textit{Kodak, 24 images}} \\
Robust Turbo-DDCM & 0.0728 & +0.48 & +0.87 & +1.87 & +3.50 & +6.36 & $-$28 & $-$42 & $-$85 & $-$166 & $-$343 & $-$21 & $-$28 & $-$49 & $-$86 & $-$164 \\
BitFlat, parity & 0.0735 & $-$0.26 & $-$0.05 & +0.46 & +1.49 & +3.78 & +21 & +14 & $-$2 & $-$41 & $-$156 & +13 & +11 & +2 & $-$18 & $-$72 \\
\addlinespace[1.5pt]
Robust Turbo-DDCM & 0.1003 & +0.54 & +0.94 & +2.13 & +3.88 & +6.91 & $-$29 & $-$42 & $-$87 & $-$170 & $-$360 & $-$17 & $-$24 & $-$48 & $-$86 & $-$175 \\
BitFlat, parity & 0.0997 & $-$0.21 & +0.03 & +0.68 & +1.81 & +4.22 & +11 & +4 & $-$16 & $-$55 & $-$166 & +5 & +2 & $-$9 & $-$29 & $-$73 \\
BitFlat, Hamming & 0.1000 & +0.06 & +0.06 & +0.08 & +0.17 & +1.35 & $-$4 & $-$4 & $-$4 & $-$7 & $-$43 & $-$3 & $-$3 & $-$3 & $-$5 & $-$22 \\
BitFlat, SECDED & 0.1003 & +0.18 & +0.18 & +0.19 & +0.26 & +0.98 & $-$11 & $-$11 & $-$11 & $-$13 & $-$34 & $-$7 & $-$7 & $-$7 & $-$8 & $-$18 \\
\addlinespace
\multicolumn{17}{@{}l}{\textit{DIV2K, 100 images}} \\
Robust Turbo-DDCM & 0.0728 & +0.56 & +0.93 & +2.03 & +3.75 & +6.66 & $-$29 & $-$39 & $-$80 & $-$156 & $-$330 & $-$20 & $-$26 & $-$45 & $-$83 & $-$167 \\
BitFlat, parity & 0.0735 & $-$0.40 & $-$0.22 & +0.39 & +1.61 & +3.99 & +24 & +19 & +1 & $-$39 & $-$141 & +18 & +15 & +5 & $-$14 & $-$62 \\
\addlinespace[1.5pt]
Robust Turbo-DDCM & 0.1003 & +0.66 & +1.08 & +2.32 & +4.24 & +7.29 & $-$28 & $-$40 & $-$79 & $-$157 & $-$339 & $-$19 & $-$26 & $-$46 & $-$83 & $-$172 \\
BitFlat, parity & 0.0997 & $-$0.30 & $-$0.08 & +0.59 & +1.88 & +4.61 & +15 & +10 & $-$8 & $-$48 & $-$158 & +11 & +8 & $-$2 & $-$22 & $-$75 \\
BitFlat, Hamming & 0.1000 & +0.10 & +0.10 & +0.12 & +0.23 & +1.29 & $-$5 & $-$5 & $-$6 & $-$9 & $-$38 & $-$5 & $-$5 & $-$5 & $-$7 & $-$23 \\
BitFlat, SECDED & 0.1003 & +0.24 & +0.24 & +0.25 & +0.31 & +1.02 & $-$12 & $-$12 & $-$12 & $-$14 & $-$34 & $-$9 & $-$9 & $-$9 & $-$10 & $-$21 \\
\bottomrule
\end{tabular}
\end{landscapetable}

\begin{landscapetable}
\centering
\caption{Sectioned codec with one BCH code minus the reported values of the
robust codecs, in the format of Table~\ref{tab:flips}.}
\label{tab:sectflips}
\TableText
\setlength{\tabcolsep}{3.4pt}
\begin{tabular}{@{}llrrrrrrrrrrrrrrr@{}}
\toprule
 & & \multicolumn{5}{c}{PSNR (dB)} & \multicolumn{5}{c}{LPIPS ($10^{-3}$)}
 & \multicolumn{5}{c}{DISTS ($10^{-3}$)} \\
\cmidrule(lr){3-7}\cmidrule(lr){8-12}\cmidrule(l){13-17}
Robust codec & bpp & 1 & 3 & 10 & 30 & 100 & 1 & 3 & 10 & 30 & 100 & 1 & 3 & 10 & 30 & 100 \\
\midrule
\multicolumn{17}{@{}l}{\textit{Kodak, 24 images}} \\
Robust Turbo-DDCM & 0.0728 & +0.84 & +1.23 & +2.23 & +3.86 & +6.72 & $-$53 & $-$67 & $-$110 & $-$191 & $-$368 & $-$38 & $-$45 & $-$66 & $-$103 & $-$181 \\
BitFlat, parity & 0.0735 & +0.10 & +0.31 & +0.82 & +1.85 & +4.14 & $-$4 & $-$11 & $-$27 & $-$66 & $-$181 & $-$4 & $-$6 & $-$15 & $-$35 & $-$89 \\
\addlinespace[1.5pt]
Robust Turbo-DDCM & 0.1003 & +0.85 & +1.25 & +2.44 & +4.19 & +7.22 & $-$45 & $-$58 & $-$103 & $-$186 & $-$376 & $-$26 & $-$33 & $-$57 & $-$95 & $-$185 \\
BitFlat, parity & 0.0997 & +0.10 & +0.34 & +0.99 & +2.12 & +4.53 & $-$5 & $-$12 & $-$32 & $-$71 & $-$182 & $-$4 & $-$7 & $-$18 & $-$38 & $-$83 \\
BitFlat, Hamming & 0.1000 & +0.37 & +0.37 & +0.39 & +0.48 & +1.65 & $-$20 & $-$20 & $-$20 & $-$23 & $-$59 & $-$12 & $-$12 & $-$12 & $-$14 & $-$31 \\
BitFlat, SECDED & 0.1003 & +0.49 & +0.49 & +0.50 & +0.57 & +1.29 & $-$27 & $-$27 & $-$27 & $-$29 & $-$50 & $-$16 & $-$16 & $-$16 & $-$17 & $-$28 \\
\addlinespace
\multicolumn{17}{@{}l}{\textit{DIV2K, 100 images}} \\
Robust Turbo-DDCM & 0.0728 & +1.07 & +1.44 & +2.54 & +4.26 & +7.18 & $-$58 & $-$68 & $-$109 & $-$185 & $-$360 & $-$40 & $-$46 & $-$65 & $-$103 & $-$187 \\
BitFlat, parity & 0.0735 & +0.08 & +0.26 & +0.87 & +2.09 & +4.48 & $-$2 & $-$7 & $-$25 & $-$65 & $-$168 & $-$1 & $-$4 & $-$14 & $-$33 & $-$82 \\
\addlinespace[1.5pt]
Robust Turbo-DDCM & 0.1003 & +1.10 & +1.52 & +2.76 & +4.68 & +7.74 & $-$47 & $-$59 & $-$98 & $-$176 & $-$357 & $-$31 & $-$38 & $-$58 & $-$95 & $-$184 \\
BitFlat, parity & 0.0997 & +0.16 & +0.38 & +1.05 & +2.34 & +5.05 & $-$5 & $-$10 & $-$28 & $-$68 & $-$176 & $-$2 & $-$5 & $-$15 & $-$35 & $-$87 \\
BitFlat, Hamming & 0.1000 & +0.54 & +0.54 & +0.56 & +0.67 & +1.73 & $-$24 & $-$24 & $-$25 & $-$28 & $-$57 & $-$17 & $-$17 & $-$17 & $-$19 & $-$34 \\
BitFlat, SECDED & 0.1003 & +0.68 & +0.68 & +0.69 & +0.75 & +1.47 & $-$31 & $-$31 & $-$31 & $-$33 & $-$52 & $-$21 & $-$21 & $-$21 & $-$22 & $-$33 \\
\bottomrule
\end{tabular}
\end{landscapetable}

Table~\ref{tab:flips} gives the results for Turbo-DDCM with one BCH code. At
26\,136 bits, the code protects 91 vectors per step in 21\,600 bits with
4\,530 parity bits, which correct up to 309 errors. The code rate is 0.83, and
the image is lost with probability $2\times10^{-3}$ at a BER of $10^{-2}$.
This baseline is better than Robust Turbo-DDCM in all three measures at every
BER. Against BitFlat with parity fields, the PSNR difference on Kodak is
$-$0.21\dB{} at $10^{-4}$ and +0.03\dB{} at $3\times10^{-4}$, and it grows to
+4.22\dB{} at $10^{-2}$. Against the Hamming and SECDED fields, the baseline
is 0.06 to 0.26\dB{} better up to $3\times10^{-3}$, which is within the
accuracy of the comparison, and 0.98 and 1.35\dB{} better at $10^{-2}$. A code
in every field spends four or five check bits on ten information bits, whereas
one code over the whole bitstream reaches a failure probability of
$3\times10^{-4}$ at a code rate of 0.82. The results on DIV2K follow the same
pattern.

Table~\ref{tab:sectflips} gives the results for the sectioned codec. On Kodak,
the sectioned bitstream has exactly the bits of the joint index that it
replaces, so the codes and their failure probabilities are unchanged and only
the quality of a received image improves. On DIV2K, two of the six designs use
the nearest available sectioned layout, which differs by 9 bits per step. All
differences in the table favour the sectioned codec.

The code must be designed for a maximum BER, and a wider range costs quality.
For Turbo-DDCM, a design for a BER of $2\times10^{-2}$ has 72 vectors per step
and a code rate of 0.68, and it is about 0.30\dB{} below the design for
$10^{-2}$. With sectioned codebooks, the design for $2\times10^{-2}$ uses 736
bits per step and a code that corrects 613 errors, and it is 0.28\dB{} below
the sectioned design for $10^{-2}$. Beyond the design range the code fails as
a whole. The design for $10^{-2}$ at 26\,136 bits loses the image with
probability 0.59 at a BER of 1.2\% and 1.00 at 1.5\%.

\section{Reproduction of Robust Turbo-DDCM}
\label{sec:flipschecks}

The comparison with BitFlat relies on reported values, so our evaluation must
match the reported one in images and quality measures. Table~I of
\cite{deng2026bitflat} includes Robust Turbo-DDCM, whose code is public. We
reproduced this codec under the reported settings, with independent bit flips
applied to its bitstream. Table~\ref{tab:calibration} lists the differences
between our PSNR values and the reported ones. They do not exceed 0.15\dB{} on
Kodak and 0.09\dB{} on DIV2K, and the differences in LPIPS and DISTS are at
most 0.005 and 0.009. Differences of this size are within the variation
between channel realizations. The agreement holds for a codec that we can run
and is not a confidence interval for the reported values of BitFlat.
Differences of one or two tenths of a decibel to those values should be
regarded as inconclusive.

\begin{table}[t]
\centering
\caption{Our reproduction of Robust Turbo-DDCM minus the reported PSNR (dB),
with 73 vectors per step (0.1003\bpp) and, on Kodak, also with 53 vectors
(0.0728\bpp).}
\label{tab:calibration}
\TableText
\setlength{\tabcolsep}{4pt}
\begin{tabular}{@{}llrrrrr@{}}
\toprule
Images & Vectors & $10^{-4}$ & $3\!\times\!10^{-4}$ & $10^{-3}$ & $3\!\times\!10^{-3}$ & $10^{-2}$ \\
\midrule
Kodak & 73 & $-$0.05 & $-$0.07 & $-$0.01 & +0.02 & $-$0.03 \\
 & 53 & +0.01 & +0.07 & +0.12 & $-$0.15 & +0.10 \\
\addlinespace
DIV2K & 73 & 0.00 & $-$0.08 & +0.07 & $-$0.07 & $-$0.09 \\
\bottomrule
\end{tabular}
\end{table}

For the third dataset of \cite{deng2026bitflat}, the CLIC 2020 professional
test set \cite{clic2020test}, we could not reproduce the reported values. On
the central crops of its 250 images, our PSNR for Robust Turbo-DDCM is
0.60\dB{} higher than the reported value at a BER of $10^{-4}$, which suggests
that the reported values were obtained on different images. We therefore do
not use the reported values for this dataset and compare the codecs in our own
evaluation. In this comparison, Turbo-DDCM with one BCH code is 0.74 to
8.91\dB{} better than Robust Turbo-DDCM, and the sectioned codec is 1.13 to
9.30\dB{} better.

\section{Burst errors}
\label{sec:bursts}

Both papers on robust diffusion codecs leave burst errors untested. We model
bursts with a two-state channel
\cite{gilbert1960capacity,elliott1963estimates}. The good state flips no bits,
and the bad state flips every bit with probability 0.5. The mean duration of
the bad state is 8, 64, 512 or 4\,096 bits, and the fraction of time spent in
it is set so that the average BER is $10^{-3}$, $3\times10^{-3}$ or $10^{-2}$.
We measured Robust Turbo-DDCM on these channels at 19\,080 bits per image and
computed the failure probabilities of the codes exactly.

\begin{widefigure}
\centering
\includegraphics{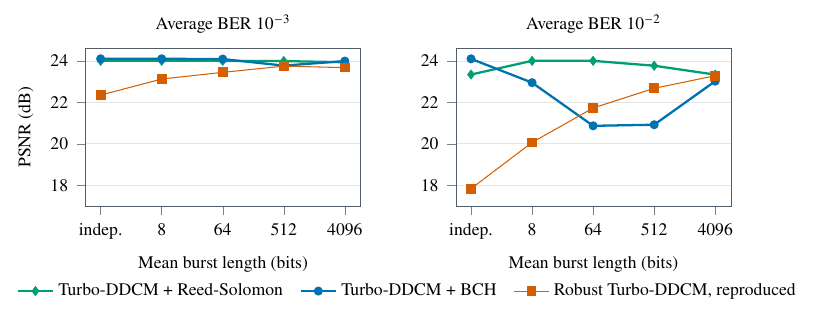}
\caption{PSNR on Kodak at 19\,080 bits per image under burst errors, as a
function of the mean burst length. The leftmost point of each panel is the
channel with independent errors. The Reed-Solomon code is designed once for
all channels of the right panel.}
\label{fig:bursts}
\end{widefigure}

A BCH code designed for independent errors is not suitable for this channel
(Fig.~\ref{fig:bursts}). At an average BER of $10^{-2}$, it loses 10.2\% of
the images with bursts of 8 bits and 28.1\% with bursts of 512 bits. A
Reed-Solomon code with symbols of 11 bits, designed once for all tested
channels at $10^{-2}$, keeps Turbo-DDCM on par with or better than Robust
Turbo-DDCM on every channel, at a cost of 0.76\dB{} when the errors are
independent. Its advantage decreases with the burst length, from 3.94\dB{} for
bursts of 8 bits to 1.09\dB{} for bursts of 512 bits at an average BER of
$10^{-2}$. For bursts of 4\,096 bits, a fifth of the bitstream, the advantage
is 0.05 to 0.44\dB, while the measured mean of the robust codec varies by 0.27
to 0.46\dB{} between realizations, so no clear advantage remains. A robust
codec benefits from burstiness, because a burst that flips many bits of one
field still corrupts only one vector, whereas a long burst can exceed the
correction capability of any code that the bit budget allows.

Sectioned codebooks add 0.36\dB{} to both coded designs wherever the code
holds. With the Reed-Solomon code, the sectioned codec is better than Robust
Turbo-DDCM on all 12 burst channels, by 0.39 to 4.30\dB{} in general and by
0.39 to 0.79\dB{} for the longest bursts.

\section{Packet loss}
\label{sec:loss}

We repeated the comparison for packet loss with a learned codec optimized for
mean squared error and an erasure code.

The robust codec is Every Packet Counts (EPC) \cite{wei2026every}, which is
trained to reconstruct the image from the latent packets that arrive, for loss
rates between 0 and 30\%. Its code and weights are not public, so we use the
reported mean PSNR at independent loss rates of 5, 10 and 20\%, on the Kodak
images at 0.125, 0.217 and 0.346\bpp{} and on images of the CLIC 2020 dataset
at 0.090, 0.148 and 0.228\bpp. EPC is built on a variant of the HPCM codec
\cite{li2025learned}. Our reference codec is MLIC++ \cite{jiang2025mlicpp}, a
codec of the same generation, so the two systems do not share a base codec.

The protected codec compresses every image with MLIC++ at one operating point
for the whole dataset and uses the actual entropy coder. The bitstream is
spread over packets of at most 1500 bytes with one maximum-distance-separable
code, and the packet headers count in the rate. Following the training range
of EPC, the design maximizes the expected PSNR for a loss rate that is uniform
between 0 and 30\%. An image is sent in at least eight packets on Kodak and at
least 32 packets on the larger images. Two differences in accounting favour
EPC. Its rates include no packet headers, and it assumes that the packet with
the hyperprior always arrives. The larger images are the 41 images of the CLIC
2020 professional validation set \cite{clic2020validation} at full size, with
2.2 megapixels on average. EPC only states that it uses the CLIC 2020 dataset
at 2K resolution, so the comparison on these images is indicative. We also
include unequal protection
\cite{albanese1996priority,mohr2000unequal,dumitrescu2004globally}, in which
the bitstream of MLIC++ is divided into at most four levels with decreasing
protection along the decoding order.

\begin{widetable}
\centering
\caption{MLIC++ protected by erasure codes designed for loss rates from 0 to
30\%, compared with the reported PSNR (dB) of Every Packet Counts at
independent loss rates of 5, 10 and 20\%. Graded denotes unequal protection.
The budget is defined in \appref{sec:budget}.}
\label{tab:loss}
\TableText
\setlength{\tabcolsep}{3.8pt}
\begin{tabular}{@{}llrrrrrrrrrrrr@{}}
\toprule
 & & \multicolumn{3}{c}{EPC, reported} & \multicolumn{3}{c}{One code minus EPC}
 & \multicolumn{3}{c}{Graded minus one code} & \multicolumn{3}{c}{Budget} \\
\cmidrule(lr){3-5}\cmidrule(lr){6-8}\cmidrule(lr){9-11}\cmidrule(l){12-14}
Images & bpp & 5\% & 10\% & 20\% & 5\% & 10\% & 20\% & 5\% & 10\% & 20\% & 5\% & 10\% & 20\% \\
\midrule
Kodak & 0.125 & 27.55 & 27.47 & 27.21 & +0.51 & +0.52 & +0.14 & +0.18 & +0.19 & +0.34 & 1.12 & 1.19 & 1.57 \\
 & 0.217 & 28.98 & 28.84 & 28.56 & +0.73 & +0.82 & +0.51 & +0.15 & +0.16 & +0.29 & 1.28 & 1.22 & 1.38 \\
 & 0.346 & 30.43 & 30.27 & 29.93 & +0.77 & +0.92 & +0.86 & +0.23 & +0.22 & +0.20 & 1.81 & 1.58 & 1.57 \\
\addlinespace
CLIC & 0.090 & 30.55 & 30.38 & 30.01 & $-$0.86 & $-$0.69 & $-$0.44 &  &  &  & 1.40 & 1.40 & 0.81 \\
 & 0.148 & 31.93 & 31.79 & 31.45 & $-$0.84 & $-$0.70 & $-$0.36 &  &  &  & 1.45 & 1.45 & 0.68 \\
 & 0.228 & 33.28 & 33.14 & 32.78 & $-$0.74 & $-$0.59 & $-$0.24 &  &  &  & 1.42 & 1.42 & 0.73 \\
\bottomrule
\end{tabular}
\end{widetable}

Table~\ref{tab:loss} shows the results. On Kodak, one erasure code is better
than EPC in all nine cases, by 0.14 to 0.92\dB, and unequal protection adds
0.15 to 0.34\dB. On the larger images the order is reversed, and one code is
0.24 to 0.86\dB{} worse than EPC. The reported values themselves differ
between the two datasets. At 5\% loss, EPC is 1.84 to 2.83\dB{} below
error-free MLIC++ at the same rate on Kodak and only 0.54 to 0.68\dB{} below
it on the larger images. A code that must survive a loss of 30\% gives up
about a third of the rate, which costs 1.33 to 2.06\dB{} on Kodak and 1.40 to
1.45\dB{} on the larger images. The two datasets differ in the images, in the
number of packets per image and possibly in the selection of images behind the
reported values. We report the reversal as an observation and do not attribute
it to the image size.

\section{The budget of a robust codec}
\label{sec:budget}

This appendix derives a limit on the gain that a robust codec can achieve over
a codec protected by a channel code, and compares the observed gains with this
limit.

Let $C(\theta)$ be the capacity of the channel per transmitted bit, which is
$1-h(p)$ for independent bit errors with the binary entropy function $h$ and
$1-q$ for packet loss. Whatever the scheme, the receiver obtains at most
$nC(\theta)$ bits of information about the image from $n$ channel uses in
state $\theta$ \cite{cover2006elements}. No scheme can therefore deliver a
better expected quality than the distortion-rate function of the images allows
at $nC(\theta)$ bits. This function is unknown, and we use the quality of the
reference codec at that number of bits instead,
\begin{equation}
G(\theta)=Q\bigl(nC(\theta)\bigr),
\label{eq:genie}
\end{equation}
\begin{equation}
B(\theta)=G(\theta)-S(\theta).
\label{eq:budget}
\end{equation}
We call $B(\theta)$ the \emph{budget} at state $\theta$. A robust codec whose
gain over the protected codec exceeds the budget delivers more than the
reference codec could deliver over an error-free channel of the same capacity.
Such a gain would be an advance in compression and not in resilience. The
budget is a yardstick and not a bound for every robust codec, because a codec
with a better rate-distortion performance than the reference codec can exceed
it.

The size of the budget follows from two numbers. If the quality of the codec
grows by $s$ decibels per doubling of the rate, and the code carries a
fraction $\rho<1$ of the capacity at $\bar\theta$, then
\begin{equation}
B(\theta)\approx s\,\log_2\frac{C(\theta)}{\rho\,C(\bar\theta)} .
\label{eq:slope}
\end{equation}
For BERs up to $10^{-2}$ the capacity decreases from 1 to 0.919, and for loss
rates up to 30\% it decreases to 0.7. With the slopes of the codecs used here,
1.1\dB{} per doubling for Turbo-DDCM and 1.7 to 2.8\dB{} for MLIC++, an ideal
code leaves a robust codec at most 0.14\dB{} at a vanishing BER and about one
decibel at a vanishing loss rate.

\begin{widefigure}
\centering
\includegraphics{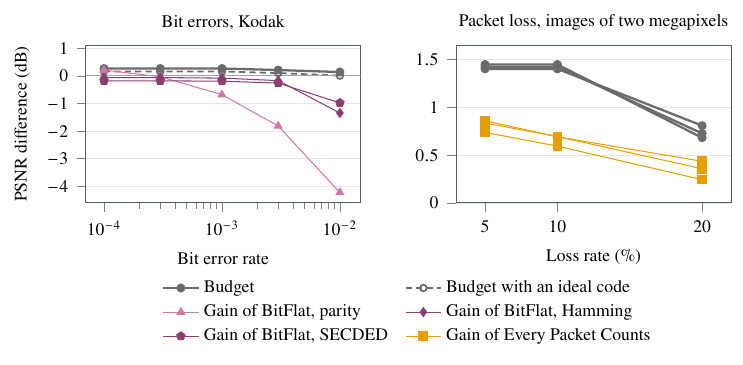}
\caption{The budget (\ref{eq:budget}) and the PSNR gains of the robust codecs
over the protected codec. Left: bit errors on Kodak at about 0.10\bpp. A
negative gain means that the protected codec is better. Right: packet loss on
images of two megapixels.}
\label{fig:budget}
\end{widefigure}

Fig.~\ref{fig:budget} compares the gains of the robust codecs with the budget.
For bit errors at 26\,136 bits on Kodak and a BER of $10^{-4}$, the budget of
Turbo-DDCM with the BCH code is 0.26\dB, and BitFlat with parity fields gains
0.21\dB, which is 78\% of the budget. At $10^{-2}$ the budget shrinks to
0.14\dB, while BitFlat with parity fields is 4.22\dB{} below the protected
codec. With the sectioned codec as the reference, the budget at $10^{-4}$ is
0.22 to 0.46\dB, and no reported value of a robust codec exceeds the protected
codec. For packet loss the budget is larger, with 1.12 to 1.81\dB{} on Kodak
and 0.68 to 1.45\dB{} on the larger images (Table~\ref{tab:loss}). EPC gains
nothing on Kodak and 34 to 61\% of the budget on the larger images. All
observed gains of the robust codecs are within the budget.

\bibliographystyle{IEEEtran}
\bibliography{references}

\end{document}